# Understanding temperature tuning in energy-based models

Peter W. Fields,[1] Vudtiwat Ngampruetikorn,[2,3] David J. Schwab,[4] and Stephanie E. Palmer[1,3,5,6,*]

[1]*Department of Physics, University of Chicago, Chicago, Illinois 60637, USA*
[2]*School of Physics, University of Sydney, Sydney, New South Wales 2006, Australia*
[3]*National Institute for Theory and Mathematics in Biology, Northwestern University and The University of Chicago, Chicago, Illinois 60611, USA*
[4]*Initiative for the Theoretical Sciences and CUNY—Princeton Center for the Physics of Biological Function, The Graduate Center, CUNY, New York, New York 10016, USA*
[5]*Department of Organismal Biology and Anatomy, University of Chicago, Chicago, Illinois 60637, USA*
[6]*Center for Living Systems, University of Chicago, Chicago, Illinois 60637, USA*

Generative models of complex systems often require *post hoc* parameter adjustments to produce useful outputs. For example, energy-based models for protein design are sampled at an artificially low "temperature" to generate novel, functional sequences. This temperature tuning is a common yet poorly understood heuristic used across machine learning contexts to control the trade-off between generative fidelity and diversity. Here, we develop an interpretable, physically motivated framework to explain this phenomenon. We demonstrate that in systems with a large "energy gap" separating a small fraction of meaningful states from a vast space of unrealistic states, learning from sparse data causes models to systematically overestimate high-energy state probabilities, a bias that lowering the sampling temperature corrects. More generally, we characterize how the optimal sampling temperature depends on the interplay between data size and the system's underlying energy landscape, and that this interplay is sufficient to elicit the tuning phenomenon; regularizing while training can contribute, but is not necessary. Crucially, our results show that lowering the sampling temperature is not always desirable. We identify the conditions under which *raising* it results in better generative performance. Our framework reveals that *post hoc* temperature tuning is indicative of both specific properties of the true data distribution and the limits of the learned model due to insufficient data availability.

## I. INTRODUCTION

Energy-based models trained on evolutionary data can now generate novel protein sequences that replicate natural function [1]. A crucial yet poorly understood step in these successes is the use of an artificially low sampling "temperature" to produce functional sequences from the trained model. This adjustment is often the deciding factor between generating functional enzymes and inert polypeptides. A fundamental question arises as to what necessitates temperature tuning and what it reveals about the space of functional proteins and the limits of the models trained on finite data.

Temperature tuning is a broadly used heuristic across machine learning contexts, used to improve training [2–4], generalization/generative performance [5–8], and energy-landscape dynamics for memory retrieval [9]. It follows the basic intuition that one can navigate the trade-off between fidelity (producing believable, high-probability outputs at low temperature) and diversity (exploring a wide range of novel outputs at high temperature). Despite its widespread use, this practice lacks a principled, quantitative explanation and has not been systematically connected to known issues of the fitting procedure—particularly how it connects to fundamental limits in the learning process, such as biases introduced by training on finite data [10–15].

Inspired by the success of energy-based models in protein synthesis, we investigate the temperature-tuning phenomenon using interpretable, physically motivated models. Our central hypothesis traces the need for temperature tuning to a common feature of high-dimensional protein systems. A vast space of noisy, improbable sequences is separated from a small set of meaningful, functional states [16]. One may conceptualize this as a large "energy gap" that separates the small region of low-energy states from the large space of high-energy ones. When trained on necessarily sparse data from such a system, a model develops a specific bias, systematically overestimating the probability of the high-energy states. Lowering the sampling temperature serves as a direct correction for this bias, suppressing the generation of infeasible output states and improving generative fidelity.

However, our framework reveals a richer and more complex picture. The optimal sampling temperature is not a fixed parameter but is determined by a quantifiable interplay between the amount of training data and the properties of the underlying energy landscape. Crucially, lowering the temperature is not always optimal. We identify the precise conditions

[*]Contact author: sepalmer@uchicago.edu

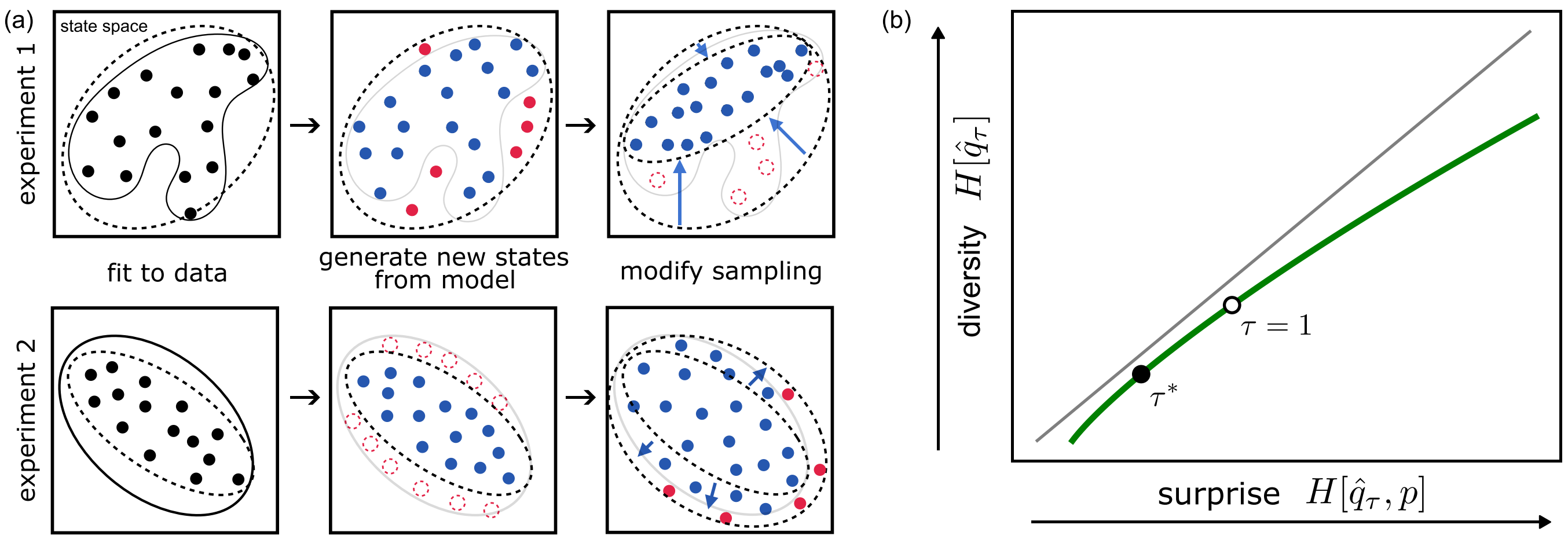

FIG. 1. (a) Schematic of training, modifying, and sampling generative models. Each box represents the entirety of state space—each dot a data point in that space. (a) (Left) Training data (black dots) are generated by a ground truth distribution, $p$ (solid line). A model is fit to these data, $\hat{q}$ (dotted lines). (a) (Middle) Generate samples from the model. Samples may either be taken from areas of high probability in the ground truth (blue dots) or from areas of low probability in the ground truth (red solid dots—false positives). Note that depending on the ground truth distribution, the fit model may also fail to generate relevant samples (red empty dots—false negatives). (a) (Right) Modifying the sampling technique gives a larger proportion of relevant samples. At top, this comes at the expense of missing some areas of ground truth distribution, increasing false negatives. At bottom, this increases the sampling of false positives. (b) The trade-off between sampling states from the fit model that are probable, i.e., less surprising/more believable, according to the true distribution vs the diversity of said states—captured by the trade-off between entropy, $H[\hat{q}_\tau]$, and cross entropy, $H[\hat{q}_\tau, p]$, where the above curve is parametrized by sampling temperature $\tau$ of the trained distribution $\hat{q}$. The difference of these two quantities is $D_{\mathrm{KL}}(\hat{q}_\tau||p)$, whose minimum (over $\tau$) is denoted by $\tau^*$. Note that $\tau^*$ need not equal 1, which would denote sampling the model "as is."

in which raising the sampling temperature maximizes generative performance.

The authors in Ref. [1] achieved *in vivo* functionality of synthetic sequences by sampling at lower temperature, such that their histogram of energies under the fitted energy function closely overlapped the histogram of the training data energies. The authors noted that a shift toward higher-energy samples when sampling at model temperature =1 was in part due to the use of regularization during training. We highlight that our experiments in Sec. II C demonstrate that regularization is not necessary to elicit the temperature-tuning effect.

To formally characterize the conditions that determine the optimal temperature, we first establish a metric that quantitatively captures the trade-off between generative fidelity and diversity. This allows us to define and investigate an optimal sampling temperature that maximizes generative performance across different systems and data regimes. We do this in systems where we know the true distribution, its energy spectrum, and its temperature. Knowing the ground truth distribution lets us compute complementary statistical distance metrics that are useful for diagnosing and interpreting the optimal adjustments to the model temperature.

## II. RESULTS

### A. Quantifying generative performance

A good generative model must balance two competing objectives: First, it must be able to create believable samples, i.e., states with high fidelity to the true data-generating process; second, these must cover as wide a variety of states as possible, beyond merely memorizing the data itself. When training data are limited, it is difficult for the fit model to satisfy both goals, and specialized sampling procedures must be adopted to achieve a trade-off between the two.

Depicted in Fig. 1(a) are two scenarios in which the sampling procedure must be modified to obtain optimal generative performance. At left, we can see a fit model, $\hat{q}$, to training data from a ground truth, $p$. In the middle, we can see that sampling such models either leads to sampling false positives, that is, states that are not probable under the ground truth (top), or false negatives, that is, missing states with significant probability mass in the ground truth (bottom). Sampling can be modified such that either (1) more believable states (according to the ground truth) are sampled (top right) at the cost of a less diverse representation or (2) a more diverse set of states is sampled at the cost of accepting unlikely false positives (bottom right).

A good metric of generative performance is one that measures this trade-off as the broadening/narrowing of sample space is conducted. Figure 1(b) introduces the quantities that track the diversity/believability properties inherent in the model sampling procedure, where $\tau$ represents the postfitting "temperature" used to modify the sampling, to be formally introduced in the next section; $\tau = 1$ represents sampling the model "as is" after fitting.

Consider taking $N$ samples from $\hat{q}_\tau$, creating a synthetic dataset $\hat{\mathcal{D}}_\tau$. The frequency with which these samples are on the states of $p$ with significant probability mass may be quantified via $\langle -\log p(v)\rangle_{v\sim\hat{\mathcal{D}}_\tau}$, where hereafter log denotes the natural logarithm. In information-theoretic terms, the quantity $-\log p(v)$ is considered the surprise of the observation $v$. The lower the average of this quantity with respect to $\hat{\mathcal{D}}_\tau$, the less surprising (or the more believable) the samples may be considered. If we take the number of samples, $N$, to infinity, we see that

$$\lim_{N\to\infty} \langle -\log p(v)\rangle_{v\sim\hat{\mathcal{D}}_\tau} = -\sum_v \hat{q}_\tau(v)\log p(v) = H[\hat{q}_\tau, p],$$

where we identify $\langle -\log p(v)\rangle_{v\sim\hat{\mathcal{D}}_\tau}$ with the cross entropy of $\hat{q}_\tau(v)$ with $p$. In contrast to cross entropy, we may consider the entropy of the fit model, $H[\hat{q}_\tau]$, as it naturally captures the variability of states *within* the model itself:

$$\lim_{N\to\infty} \langle -\log \hat{q}_\tau(v)\rangle_{v\sim\hat{\mathcal{D}}_\tau} = -\sum_v \hat{q}_\tau(v)\log \hat{q}_\tau(v) = H[\hat{q}_\tau].$$

We note that the difference of these two quantities,

$$D_{\mathrm{KL}}(\hat{q}_\tau||p) = H[\hat{q}_\tau, p] - H[\hat{q}_\tau], \tag{1}$$

is simply the Kullback-Leibler divergence of $\hat{q}_\tau$ with $p$, giving us the desired metric that captures this believability/diversity trade-off. It has been noted elsewhere that including $D_{\mathrm{KL}}(q||p)$ (or a proxy for it) in the objective function causes the fit model to focus more strongly on modes of the data distribution, as opposed to using $D_{\mathrm{KL}}(p||q)$ alone, and is less susceptible to assigning probability mass to spurious states [17–20], leading to improved generative performance. This follows from the asymmetry of the KL divergence. Since for $D_{\mathrm{KL}}(p||\hat{q}) = \langle \log(p/\hat{q})\rangle_p$, the average is taken with respect to $p$, any place where $p$ is appreciable but $\hat{q}$ is small, the $D_{\mathrm{KL}}$ will be heavily penalized, forcing $\hat{q}$ to be spread out over the full support. For $D_{\mathrm{KL}}(\hat{q}||p) = \langle \log(\hat{q}/p)\rangle_{\hat{q}}$, it is penalized where $p$ is small, and $\hat{q}$ is forced to focus on areas where the mass is high in $p$ and abandon its tails.

We make use of $D_{\mathrm{KL}}(\hat{q}_\tau||p)$, which has been called the *reversed* $D_{\mathrm{KL}}$ [21], as it reverses the arguments of the typical objective function used for fitting in many machine learning contexts. We refer to the usual distance metric, $D_{\mathrm{KL}}(p||q)$, as the *forward* $D_{\mathrm{KL}}$.

As shown in Fig. 1(b), as we tune the sampling temperature, $\tau$, obtaining probable states comes at the expense of diversity (lower entropy); having less surprising samples implies less diverse samples, and vice versa. Furthermore, the minimum of the difference of $H[\hat{q}_\tau, p]$ and $H[\hat{q}_\tau]$ at $\tau^*$ corresponds to the minimum of Eq. (1) with respect to $\tau$. In our experiments on the Ising distribution, this minimum is empirically found to be unique.

To summarize, $D_{\mathrm{KL}}(\hat{q}_\tau||p)$ (the reversed $D_{\mathrm{KL}}$) captures the trade-off between the ability of a fit model to generate diverse samples versus probable samples with respect to the ground truth distribution, $p$. This motivates the existence of an optimal sampling temperature, $\tau^*$, that produces states representing the best trade-off between the two.

### B. An illustrative model

To illustrate what necessitates tuning model temperature to improve generative performance, we set up a simple experiment on a toy model, as depicted in Fig. 2. The ground truth from which we draw samples is given by

$$p_i = \frac{\exp(-\Delta L_i)}{Z(\Delta, \mathbf{L})}, \tag{2}$$

where $\mathbf{L}$ is a vector that assigns each state $i$ to a low- or high-energy level, $L_i \in \{0, 1\}$, $\Delta$ is the energy gap between levels, and $Z(\mathbf{L}, \Delta) = n_l + n_h \exp(-\Delta)$ is the partition function, where $n_l$ and $n_h$ are the number of low- and high-energy states. Sampling this model gives the empirical distribution over states, represented as a vector, $\mathbf{p}_{\mathrm{data}}$. The goal of learning is to get best estimates of the level assignment vector, $\hat{\mathbf{L}}$, and the energy gap, $\hat{\Delta}$. Shown in Fig. 2(a) is the workflow of defining a ground truth, generating training samples, and fitting.

This toy model has desirable properties analogous to those in real learning settings, especially when considering the case, where $\frac{n_h}{n_l} \gg 1$ and $\Delta \to \infty$ [22]. In this case, low-energy states may be thought of as the "realistic" samples within a dataset and high-energy states as "unrealistic," e.g., the small number of viable proteins relative to the large number of sequences in the support that the distribution (theoretically) allows for. Importantly, fitting $\hat{\mathbf{L}}$ and $\hat{\Delta}$ corresponds to discovering meaningful states and adequately segregating them from meaningless states. The value of $\hat{\Delta}$ controls the degree of this separation and directly controls how often the fit model will generate "realistic" states when sampled.

As is typically done with real data, the parameters that minimize $D_{\mathrm{KL}}(\mathbf{p}||\mathbf{q})$, where $\mathbf{q}$ represents our model ansatz with the same functional form as $\mathbf{p}$, are found via an empirical approximation [23] to the true distribution using $\mathbf{p}_{\mathrm{data}}$:

$$\begin{aligned} D_{\mathrm{KL}}(\mathbf{p}||\mathbf{q}_{\tilde{\Delta},\tilde{\mathbf{L}}}) &\approx D_{\mathrm{KL}}(\mathbf{p}_{\mathrm{data}}||\mathbf{q}_{\tilde{\Delta},\tilde{\mathbf{L}}}) \\ &= -\sum_i p_{\mathrm{data},i}\log q_{i|\tilde{\Delta},\tilde{\mathbf{L}}} - H[\mathbf{p}_{\mathrm{data}}] \\ &= -\mathcal{L}(\tilde{\Delta}, \tilde{\mathbf{L}}) + \text{constant}, \end{aligned} \tag{3}$$

defining the loss function used to get best estimate parameters,

$$\hat{\Delta}, \hat{\mathbf{L}} = \underset{\tilde{\Delta},\tilde{\mathbf{L}}}{\operatorname{argmax}}\ \mathcal{L}(\tilde{\Delta}, \tilde{\mathbf{L}}), \tag{4}$$

with

$$-\mathcal{L}(\tilde{\Delta}, \tilde{\mathbf{L}}) = \tilde{\Delta}\tilde{\mathbf{L}}\cdot\mathbf{p}_{\mathrm{data}} + \log Z(\tilde{\Delta}, \tilde{\mathbf{L}}), \tag{5}$$

which correspond to the maximum likelihood estimates as well, where $-\mathcal{L}(\tilde{\Delta}, \tilde{\mathbf{L}})$ is the negative log-likelihood function [24].

For a given estimate of $\hat{\mathbf{L}}$, the estimate of the energy gap is given by

$$\hat{\Delta} = \log\left(\frac{\hat{n}_h}{\hat{n}_l}\cdot\frac{1-\hat{\mathbf{L}}\cdot\mathbf{p}_{\mathrm{data}}}{\hat{\mathbf{L}}\cdot\mathbf{p}_{\mathrm{data}}}\right), \tag{6}$$

where $\hat{n}_h = \sum_i \hat{L}_i$ and $\hat{n}_l = N_s - \sum_i \hat{L}_i$ are the model estimates for number of high- and low-energy states; $N_s = n_h + n_l = \hat{n}_h + \hat{n}_l$ is the total number of states and is fixed. (For reference, for a model of a protein of sequence length = 100 and ∼20 possible amino acids that can occupy each position, the model's support has a total of $N_s = 20^{100}$ possible states.)

This minimal model illustrates the interplay between the training sample size $M$ and the ground truth energy gap $\Delta$ that leads to the generative temperature-tuning effect. We are interested in the regime where high-energy "meaningless" states outnumber low-energy "meaningful" states, and training data number is low, resolving this difference poorly. A simple instance of this, where $\Delta = 5$, $n_l = 3$, $n_h = 7$, and $M = 5$ samples, giving the empirical distribution, $\mathbf{p}_{\mathrm{data}}$, is shown in Fig. 2(b). For undersampled training sets such as this, states 3 to 10 have not been sampled. Consequently, states 1 and 2 are assigned as low-energy states by the model and the objective function minimum "misses" one of the low-energy states, assigning it as high-energy, as highlighted in Fig. 2(a).

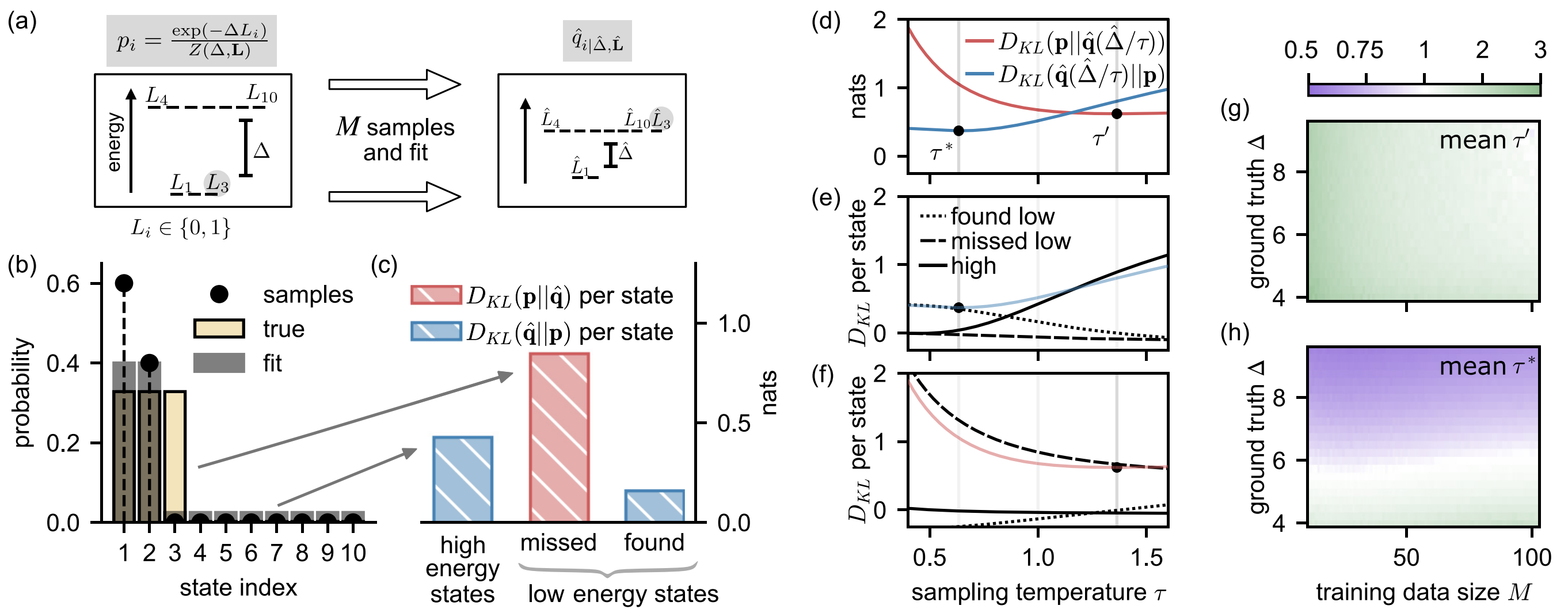

FIG. 2. Simple toy example of when raising vs lowering temperature improves generative performance. (a) The ground truth consists of a vector, $\mathbf{L}$, that assigns each of ten states to a low- or high-energy level. Sampling this distribution leads to an empirical distribution over states, $\mathbf{p}_{\mathrm{data}}$, used to get maximum likelihood estimates of the energy level assignment vector and energy gap, $\hat{\mathbf{L}}$ and $\hat{\Delta}$. (b) True model in yellow with $\Delta = 5$ and data distribution, made from five samples, represented by dotted lines. Fitting to an undersampled distribution causes the maximum likelihood estimates to overestimate the probability mass on high-energy states and underestimate the mass on the "missed" low-energy state. (c) Checking the decomposition of the forward (red) and reversed (blue) $D_{\mathrm{KL}}$'s between the fit and true distributions. Note the main contributions to each $D_{\mathrm{KL}}$: the missed low-energy states for the forward and the high-energy states for the reverse. (d) Rescaling the inferred energy gap by $\tau$, while keeping $\hat{\mathbf{L}}$ fixed, affects forward and reversed $D_{\mathrm{KL}}$'s. Note that temperature must be changed in opposite directions to achieve minima for each. The minimum of the reversed (blue) $D_{\mathrm{KL}}$ corresponds to $\tau^*$ in Fig. 1. (e), (f) $D_{\mathrm{KL}}$'s decomposed into contributions from missed low-energy states, found low-energy states, and high-energy states. Note that raising the temperature, $\tau$, leads to mitigation of the contribution from the missed low-energy state for the forward $D_{\mathrm{KL}}$ in panel (f) and lowering $\tau$ leads to mitigation of the contribution from the high-energy states to the reversed $D_{\mathrm{KL}}$ in panel (e). (g), (h) Scaled color images of mean optimal $\tau^*$ and $\tau'$—calculated from 50 replicates on experiments of a ground truth with $n_l = 20$ and $n_h = 80$ for several values of $M$ and $\Delta$ (see Appendix A 3, for details regarding calculation of $\tau^*$ and $\tau'$).

In Fig. 2(c), the performance of the fit model $\hat{q}$ is measured according to the two Kullback-Leibler divergences $D_{\mathrm{KL}}(\mathbf{p}||\hat{\mathbf{q}})$ and $D_{\mathrm{KL}}(\hat{\mathbf{q}}||\mathbf{p})$, the forward and reversed $D_{\mathrm{KL}}$, respectively. Each of these quantities is decomposed into the sum over the different states, i.e.,

$$D_{\mathrm{KL}}(f||g) = \sum_i f_i \log \frac{f_i}{g_i}$$
$$= \sum_{\substack{i \in \text{found} \\ \text{low}}} f_i \log \frac{f_i}{g_i} + \sum_{\substack{i \in \text{missed} \\ \text{low}}} f_i \log \frac{f_i}{g_i} + \cdots,$$

where the sum continues over all other states.

It is clear that the forward $D_{\mathrm{KL}}$ (red) is severely penalized by the missed low-energy state, whereas the reversed $D_{\mathrm{KL}}$ (blue) suffers more from contributions of the high-energy states. Note that while each high-energy state has low probability in the fit model, the larger number of such states, when compared to the number of low-energy states, leads to a large penalty in the reversed $D_{\mathrm{KL}}$. In addition, the reversed $D_{\mathrm{KL}}$ highlights the deleterious effects of higher-energy states—the states that are not desirable for generative performance—where the forward $D_{\mathrm{KL}}$ cannot. The forward $D_{\mathrm{KL}}$ is exactly the function our objective approximated [Eq. (3)], and its failure to capture the harm to generativity reflects the misalignment of learning objective and desired performance (see Appendix D for further details of how this misalignment contributes to the temperature-tuning effect).

The generative performance of our fit model is interrogated via rescaling $\hat{\Delta}$ by $\tau$ while maintaining the estimate of $\hat{\mathbf{L}}$. Figure 2(d) shows the effect on each $D_{\mathrm{KL}}$. The minimum value for the forward $D_{\mathrm{KL}}$ is achieved by raising the temperature. This follows the intuition that an undersampled fit with low entropy ought to have its temperature raised in order to increase probability of those states missed. Figure 2(f) shows this explicitly, as this missed state's contribution to the forward $D_{\mathrm{KL}}$ is the main one mitigated by raising temperature.

Lowering the temperature mitigates the contribution to the reversed $D_{\mathrm{KL}}$ from the high-energy states, as shown in Fig. 2(e). This corresponds well with the intuition built up from Fig. 1: Lowering the temperature allows for sampling more viable states at the cost of lower diversity.

Figures 2(g) and 2(h) depict the results of several experiments on a larger system with $n_l = 20$, $n_h = 80$ for several values of $\Delta$ and number of training data, $M$. From Fig. 2(h), we can see the tendency for low $M$ and high $\Delta$ to lead generically to the need to lower $\tau$ in order to achieve $\tau^*$. Furthermore, we can identify the regime where $\Delta$ and $M$ are small as the conditions under which it becomes favorable to *raise* $\tau$ (cf. Appendix C for more details). Note also that when considering optimal $\tau$ according to the forward $D_{\mathrm{KL}}$, it is never beneficial to lower $\tau$, as shown in Fig. 2(g).

The basic intuition from this simplified setting suggests that an undersampled dataset with a wide energy gap between its low- and high-energy states necessitates lowering the temperature on a fit model. Additionally, the comparatively larger number of high-energy states determines the extent to which generative performance is harmed, and therefore, the extent to which $\tau$ must be lowered. In short, systems with a large energy gap and few samples will generally benefit from lower temperature sampling, while systems with small gaps and few samples may benefit from raising the temperature; abundant data means temperature need not be modified.

We expect these intuitions to hold beyond this toy model setting, as many true data-generating distributions should exhibit a similar tendency to have far more "meaningless" high-energy states than "meaningful" low-energy ones. The available datasets from these systems will typically be undersampled.

### C. Structured energy landscape

To test whether the intuitions from our simple two-level illustration carry to more general settings, we conduct experiments on training samples drawn from a nearest-neighbor Ising model.

The ground truth distribution from which we draw our training data is defined as

$$p_T(v) = \frac{1}{Z(T)} \exp\left(\frac{1}{T}\sum_{\langle ij\rangle} J^0 v_i v_j\right), \tag{7}$$

where $\langle ij\rangle$ indicates the sum is taken over all nearest neighbors on the two-dimensional $4\times 4$ lattice with periodic boundary conditions and $J^0 = 1$. We take $M$ samples from this at temperature $T$ to make the dataset $\mathcal{D}_T = \{v^{(1)}, v^{(2)}, \ldots, v^{(M)}\}$. We then fit to these samples by minimizing the negative log-likelihood objective function via gradient descent (see Appendix B 1) such that $\hat{\mathbf{J}} = \operatorname{argmin}_{\tilde{\mathbf{J}}}(-\mathcal{L}(\mathcal{D}_T|\tilde{\mathbf{J}}))$ given by

$$-\mathcal{L}(\mathcal{D}_T|\tilde{\mathbf{J}}) = -\frac{1}{M}\sum_{m=1}^{M} \log q(v^{(m)}|\tilde{\mathbf{J}}), \tag{8}$$

where the model ansatz is given by

$$q(v \mid \tilde{\mathbf{J}}) \propto \exp[-E(v \mid \tilde{\mathbf{J}})], \tag{9}$$

and the energy function is defined as

$$E(v \mid \tilde{\mathbf{J}}) = -\sum_{i<j} \tilde{J}_{ij} v_i v_j. \tag{10}$$

We then take the fit-model energy function and rescale it by a posttraining temperature, $\tau$,

$$\hat{q}(v|\hat{\mathbf{J}}, \tau) = \exp[-E(v|\hat{\mathbf{J}})/\tau]/\hat{Z}(\tau). \tag{11}$$

The performance of samples drawn from $\hat{q}$ may then be measured by $D_{\mathrm{KL}}(p_T||\hat{q}_\tau)$ and $D_{\mathrm{KL}}(\hat{q}_\tau||p_T)$. This workflow is depicted in Fig. 3(a). Note that the ansatz for our fit model does not know *a priori*, $J_{ij}$'s are nonzero in the ground truth; both the spatial structure and coupling strength are inferred from data.

Each $D_{\mathrm{KL}}$ may be broken down by the contribution from each state. A natural way to do this is to consider those states, $v$, that are on the same energy level in the ground truth distribution, that is,

$$\Lambda_i = \{v_k : E(v_k|\mathbf{J}^0) = E_i\}, \tag{12}$$

where $E_i$ is the energy of the ground truth distribution's $i$th level, with $E_0$ the ground state energy and increasing $i$ are higher and higher energies. We then consider $D_{\mathrm{KL}}(f||g)$ as

$$\begin{aligned} D_{\mathrm{KL}}(f||g) &= \sum_{v\in\Lambda_0} f(v)\log\frac{f(v)}{g(v)} + \sum_{v\in\Lambda_1} f(v)\log\frac{f(v)}{g(v)} \\ &\quad + \sum_{v\in\Lambda_2} f(v)\log\frac{f(v)}{g(v)} + \cdots \\ &= \sum_i \sum_{v\in\Lambda_i} f(v)\log\frac{f(v)}{g(v)} \\ &= \sum_i d_{\mathrm{KL}}(f||g)_{\Lambda_i}, \end{aligned} \tag{13}$$

where we have defined $d_{\mathrm{KL}}(f||g)_{\Lambda_i}$ to be the per-energy-level $D_{\mathrm{KL}}$.

Furthermore, to compare the total amount of probability mass on each $\Lambda_i$ assigned by each model, we consider the quantities

$$\hat{q}_i = \frac{1}{g_i}\sum_{v\in\Lambda_i} \hat{q}(v), \tag{14}$$

$$p_{T,i} = \frac{1}{g_i}\sum_{v\in\Lambda_i} p_T(v) = \frac{1}{Z(T)}\exp(-E_i/T), \tag{15}$$

where $g_i = |\Lambda_i|$ is the density of states per energy level.

By tracking the average probability per state in each energy level, as defined by Eqs. (14) and (15), we see which kinds of states are typically overestimated or underestimated by the fit model. In Fig. 3(c), the associated probabilities per state of $\hat{q}$ and $p_T$ in each energy level are shown for one experiment at low $T$ and $M$. Note that lower excited states are, on average, underestimated, $p_{T,i} > \hat{q}_i$ for $i = 1, 2$, and 3, while higher excited states are overestimated, $p_{T,i} < \hat{q}_i$, for $i = 4$ and up.

We expect that higher-energy excited states, when overestimated by the fit model, have a greater contribution to the reversed $D_{\mathrm{KL}}$, thereby representing a harmful effect on generative performance. Figure 3(d) shows the breakdown of each $D_{\mathrm{KL}}$ in accordance with Eqs. (12) and (13). Lower-energy states in the nearest-neighbor Ising model (excited states 1 through 3, which are underestimated by $\hat{q}$) form the main contribution to $D_{\mathrm{KL}}(p_T||\hat{q})$. Overestimation of higher excited states by $\hat{q}$ gives the main contribution to $D_{\mathrm{KL}}(\hat{q}||p_T)$. This is in spite of the fact that the fitted probability per state is small [Fig. 3(c)]. The larger number of states in higher-energy levels [cf. Fig. 3(b)] greatly multiplies the deleterious effects of each $\hat{q}(v)$ to $D_{\mathrm{KL}}(\hat{q}||p_T)$, as can be seen in Figs. 2(b) and 2(c) as well.

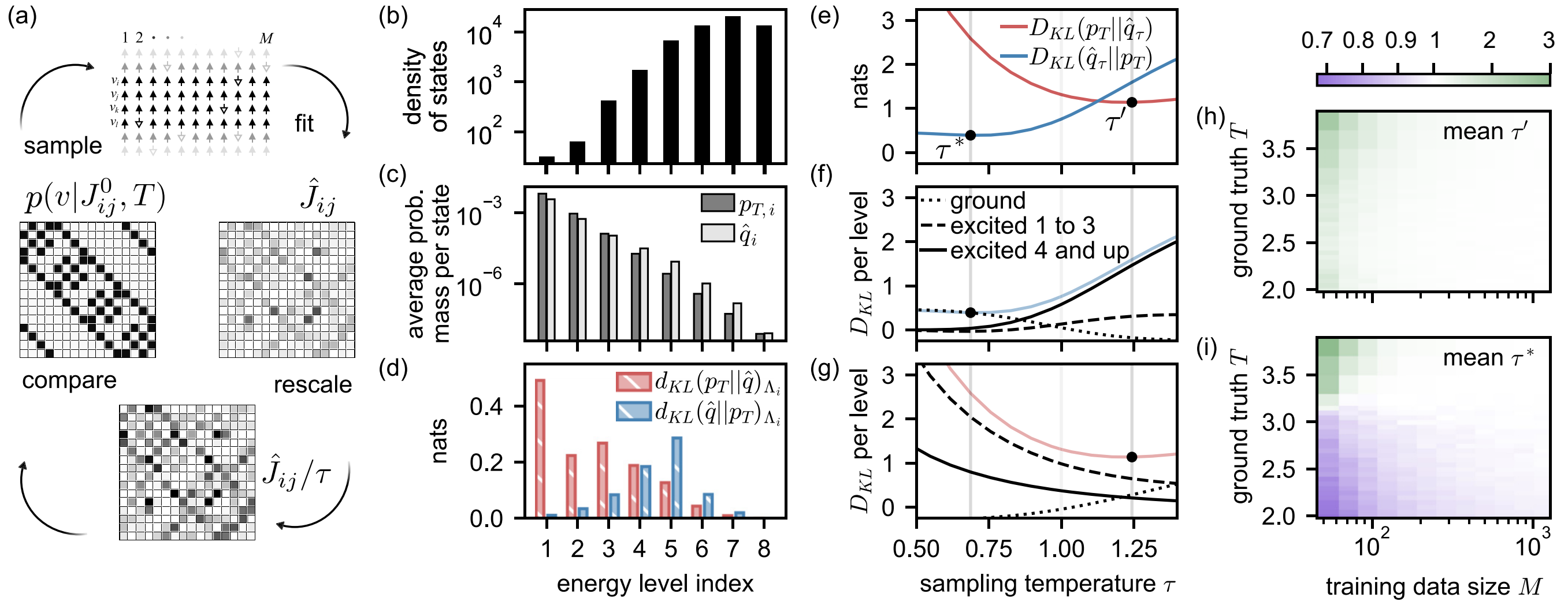

FIG. 3. Experiments on a $4 \times 4$ nearest-neighbor Ising model. (a) Starting at left and going clockwise, the ground truth model is defined at a given temperature $T$ by Eq. (7). $M$ samples are taken to form the training set, $\mathcal{D}_T$. These data are fit via minimization of Eq. (8) to give the parameters $\hat{\mathbf{J}}$. Grayscale matrices represent connectivity between different spins, with darker squares representing stronger connections. The generative properties are measured via $D_{\mathrm{KL}}(p_T||\hat{q}_\tau)$ and $D_{\mathrm{KL}}(\hat{q}_\tau||p_T)$ (see text). (b) Density of states for the first eight excited energy levels of a $4 \times 4$ nearest-neighbor Ising model. (c)–(g) Results from one experiment, where $M = 71$ and $T = 2.05$. (c) The average amount of probability per state within each energy level, as described by Eqs. (14) and (15). The amount of probability per state is underestimated by $\hat{q}$ for lower excited states and overestimated for higher excited states. (d) The contribution to each $D_{\mathrm{KL}}$ per energy level. The lower excited states are more deleterious to the forward $D_{\mathrm{KL}}$ (red) and the higher excited states are more deleterious to the reversed $D_{\mathrm{KL}}$ (blue). (e)–(g) Raising vs lowering sampling temperature $\tau$ and its dependence on contributions to each $D_{\mathrm{KL}}$ from states at different energy levels. (e) $D_{\mathrm{KL}}$'s as a function of sampling temperature $\tau$. Note the minima of each are located on opposite sides of $\tau = 1$. (f) The reversed $D_{\mathrm{KL}}$ per energy level. Lowering $\tau$ to $\tau^*$ mainly mitigates contributions from higher excited states. (g) The forward $D_{\mathrm{KL}}$ broken down by contributions from different energy levels. Raising $\tau$ to $\tau'$ decreases the major contribution from excited states 1–3. (h), (i) Ten replicates of an experiment at each $T$ and $M$ are conducted and the corresponding optimal $\tau^*$ and $\tau'$ are found for each. The scaled color image depicts the average over replicates.

In Figs. 3(e)–3(g), we see how optimal sampling temperature mitigates different pathologies captured by the different $D_{\mathrm{KL}}$'s [cf. Figs. 2(d)–2(f)]. Sampling temperature $\tau$ must be changed in opposite directions to improve performance, depending on choice of $D_{\mathrm{KL}}$ [Fig. 3(e)]. The contributions from overestimated excited states must be mitigated by a lower $\tau$ to lower the reversed $D_{\mathrm{KL}}$ [panel (f)]. Raising $\tau$ mitigates the contributions from "missed" lower-energy states to the forward $D_{\mathrm{KL}}$ [panel (g)].

With regularity, at low $T$ and low $M$, $\tau$ must be lowered to $\tau^*$ in order to improve $D_{\mathrm{KL}}(\hat{q}_\tau||p_T)$, as shown in Fig. 3(i). Additionally, we can see the regime in which $\tau$ must be *raised* in order to improve generative performance [25]. Ten replicates of each experiment were done for a ground truth distribution of Eq. (7) at several different values of $T$ and $M$. The average value of $\tau^*$ over each set of replicates is reported, generically showing the requirement to lower $\tau$ at low training sample number, $M$, and low ground truth $T$. Interestingly, $\tau'$, which minimizes the forward $D_{\mathrm{KL}}$, must always be raised from $\tau = 1$, if at all [Fig. 3(h)], distinguishing $\tau^*$s and the reversed $D_{\mathrm{KL}}$'s ability to capture the necessity for lowering sampling temperature. Finally, we emphasize that the models were trained without regularization, showing that it is not necessary to induce the temperature-tuning effect.

## III. DISCUSSION

We have established a strong and quantitative connection between the need to lower sampling temperature, the amount of available training data, and important properties of the true data-generating distribution—namely, the size of meaningful state space versus total state space and the energy landscape that segregates these states, captured by a ground truth density of states $g_i$ and ground truth temperature $T$. These quantities are especially interesting because learning them is representative of the fundamental goals of learning a generative model: to learn the underlying structure of the data, be able to generalize to states not seen, and determine their relevance for desired performance.

In fact, in many learning contexts the high-dimensional state space is vastly undersampled, and only some small fraction of this state space has functional states. The fact that a small fraction of the state space has most of the probability mass is analogous to a system with a wide energy gap between its ground and excited states (or low *true* temperature that effectively increases the gap). The need to lower *sampling* temperature in fit models arises from the fact that the model inaccurately overestimates the probability mass on excited states. Furthermore, we have shown that this bias, and the need to mitigate it via temperature tuning, is not entirely due to regularization of the objective function during training.

Other generative modeling techniques beyond energy-based methods are currently employed for protein design. In Ref. [26], the authors indicated the need for low-temperature sampling in their diffusion model, leading to better protein backbones at the cost of lower conformational diversity, similar to our fidelity-diversity trade-off. With the autoregressive protein language model ProGen [27], the authors used top-$p$ sampling to truncate the tails of softmax next-token distributions; temperature parameters in ESM3 [28] were similarly used to control the sharpness of the softmax distribution. In our framework, the data-sparse limit, where specialized sampling is warranted, occurs when $M \ll n_l \exp(\Delta/T)$, as outlined in Appendix A 4. Whether or not other generative models for proteins are in an analogous data-sparse regime warrants further investigation, but the need for specialized sampling techniques in these other models is consistent with an analogous bias mechanism.

The temperature-tuning phenomenon in energy-based models may have implications beyond applications in protein science, as these models are used in various fields across biology, where similar conditions (low sample number and small meaningful subspace) hold. The maximum entropy principle [29], a form of energy-based modeling whereby a distribution is fit to data such that it reproduces observed statistics but is otherwise as uncertain as possible, has seen success in discovering underlying principles across several biological domains: (1) within the context of data-driven protein modeling [14,30–32], (2) ecology [33], (3) collective behavior in animal groups [34,35], (4) cell regulatory [36] and signaling networks [37], and (5) theoretical neuroscience [38–42]. It would be especially interesting to see if *raising* the temperature may improve performance of models in these various biological contexts, as our framework predicts this is a viable possibility.

Energy-based models belong to a wider class of machine learning techniques, known as generative modeling [20,43–48], which aims to learn a probability distribution of a data-generating process given samples and have been used in scientific research as tools for discovery of principles underlying complex, high-dimensional systems [49–56].

Since our framework makes the connection between optimal sampling temperature and properties of the ground truth explicit, it opens up the possibility to investigate how this optimum may scale with the true number of ground states, system size, the true distribution's temperature, and number of training data. The analysis could be extended to ground truth energy landscapes beyond the nearest-neighbor Ising model. Knowing generic relationships that hold across system size and data-generating distributions could improve training and performance across a wide variety of systems; such relationships could constrain optimal sampling temperatures in lieu of knowledge of the ground truth. We emphasize, however, that while temperature tuning may be a viable approach to fit-model correction, it may also represent a limit on the extent to which a model can be improved if additional data remain unavailable. The extent to which *post hoc* tuning may improve generative performance and the extent to which it may diagnose model failure (only solvable by more data) remain questions for future research.

Our work combines concepts from statistical physics and machine learning to explain why generative models often require *post hoc* correction. We investigate the interplay between the physical structure of the ground truth, statistical biases from finite data, and the behavior of performance metrics, explaining the necessity of temperature tuning. This perspective is particularly relevant for biological systems, such as the space of functional protein sequences, where the optimal sampling temperature can become a probe for model performance and its limitations due to sparse data. Our work provides a foundation for designing more robust training objectives and for using temperature tuning to quantitatively assess how well a model has learned the essential features of a complex biological distribution.

## ACKNOWLEDGMENTS

We thank Kyle Bojanek, Cheyne Weis, and Adam Kline for helpful comments. This work was supported by the Physics Frontier Center for Living Systems through the National Science Foundation Award No. NSF PHY-2317138; the NSF-Simons National Institute for Theory and Mathematics in Biology, Award No. NSF DMS-2235451 and Simons Foundation MP-TMPS-00005320; and the University of Chicago Materials Research Science and Engineering Center Award No. NSF DMR-2011854; the Center for the Physics of Biological Function, NSF PHY-1734030; the University of Chicago Center for Physics of Evolving Systems. Additional support came from the Polymaths Program from Schmidt Sciences, LLC, to S.E.P. and by a grant from ICAM, the Institute for Complex Adaptive Matter, to P.W.F.

## DATA AVAILABILITY

The model data and code that support the findings of this article are openly available at Ref. [57].

## APPENDIX A: SIMPLE TOY MODEL

### 1. Description

As described in Sec. II B, the probability distribution of the simple toy model is given by Eq. (2), reproduced here

$$p_i = \frac{\exp(-\Delta L_i)}{Z(\Delta, \mathbf{L})},\quad E_i = \Delta L_i, \tag{A1}$$

where $\mathbf{L}$ is a vector that assigns each state $i$ to a low- or high-energy level, $L_i \in \{0, 1\}$, $\Delta$ is the energy gap between levels, and $Z(\mathbf{L}, \Delta) = n_l + n_h \exp(-\Delta)$ is the partition function, where $n_l$ and $n_h$ are the number of low- and high-energy states.

The goal of inference is to find best estimates of the level assignment vector, $\hat{\mathbf{L}}$ and energy gap between levels, $\hat{\Delta}$. The objective function to be minimized, corresponding to maximum likelihood estimation, is

$$-\mathcal{L}(\tilde{\Delta}, \tilde{\mathbf{L}}) = \tilde{\Delta}\tilde{\mathbf{L}} \cdot \mathbf{p}_{\mathcal{D}} + \log Z(\tilde{\Delta}, \tilde{\mathbf{L}}), \tag{A2}$$

where the best estimates of model parameters are

$$\hat{\Delta}, \hat{\mathbf{L}} = \underset{\tilde{\Delta}, \tilde{\mathbf{L}}}{\operatorname{argmin}}[-\mathcal{L}(\tilde{\Delta}, \tilde{\mathbf{L}})], \tag{A3}$$

and for a given $\tilde{\mathbf{L}}$, we have

$$\tilde{\Delta} = \log \frac{\tilde{n}_h(1 - \tilde{\mathbf{L}} \cdot \mathbf{p}_{\mathcal{D}})}{\tilde{n}_l(\tilde{\mathbf{L}} \cdot \mathbf{p}_{\mathcal{D}})}, \tag{A4}$$

given by Eqs. (5) and (6), respectively.

Sampling states from Eq. (A1) gives the empirical distribution, $\mathbf{p}_{\mathcal{D}}$, where $p_{\mathcal{D},i} \in [0, 1]$ is the measured frequency of state $i$ from $M$ samples.

Note that the objective function [Eq. (A2)] also corresponds to a data approximation of $D_{\mathrm{KL}}(p_i||q_{i|\hat{\Delta},\hat{\mathbf{L}}})$ (up to an additive constant that does not affect inference) given by $D_{\mathrm{KL}}(\mathbf{p}_{\mathcal{D}}||\mathbf{q}_{\hat{\Delta},\hat{\mathbf{L}}})$.

Some properties of $\hat{\mathbf{L}} \cdot \mathbf{p}_{\mathcal{D}}$ are worth noting. If we denote the set of states labeled excited by the fit model as $\hat{e} = \{i_{\hat{e}}\}$ and recall that $\hat{L}_i = 0$ for all ground states, then we can see that

$$\hat{\mathbf{L}} \cdot \mathbf{p}_{\mathcal{D}} = \sum_i \hat{L}_i p_{i,\mathcal{D}} = \sum_{i\in\hat{e}} p_{i,\mathcal{D}}. \tag{A5}$$

The quantity $\tilde{\Delta}\tilde{\mathbf{L}} \cdot \mathbf{p}_{\mathcal{D}}$ may be thought of as the energy of states averaged over the data distribution:

$$\tilde{\Delta}\tilde{\mathbf{L}} \cdot \mathbf{p}_{\mathcal{D}} = \sum_i \tilde{\Delta}\tilde{L}_i p_{i,\mathcal{D}} = \sum_i \tilde{E}_i p_{i,\mathcal{D}} = \langle \tilde{E}_i \rangle_{\mathcal{D}}. \tag{A6}$$

Furthermore, for a well-sampled distribution we expect to fit such that true parameters are recovered: $\hat{\mathbf{L}} = \mathbf{L}$, $p_{i,\mathcal{D}} = p_i$ and therefore,

$$\lim_{M\to\infty} \hat{\mathbf{L}} \cdot \mathbf{p}_{\mathcal{D}} = \sum_i L_i p_i = \sum_{i\in\{i_e\}} p_i = \frac{n_h \exp(-\Delta)}{n_l + n_h \exp(-\Delta)}. \tag{A7}$$

### 2. Fitting procedure and regularization

We briefly note here that for undersampled datasets in this setting it is possible for $\hat{\mathbf{L}} \cdot \mathbf{p}_{\mathrm{data}} = 0$, and therefore, $\mathcal{L}(\hat{\Delta}, \hat{\mathbf{L}}) = \log \frac{n_l}{N_s}$ and $\hat{\Delta} \to \infty$. To avoid such a divergence, and to ensure a fit model with support on all states, we introduce a pseudocount regularization that assumes the probability of seeing a given high-energy state under the model is $\approx 1/(M + 1)$ if and only if $\hat{\mathbf{L}} \cdot \mathbf{p}_{\mathrm{data}} = 0$,

$$\frac{1}{M+1} = \frac{\hat{n}_h \exp(-\hat{\Delta})}{\hat{n}_l + \hat{n}_h \exp(-\hat{\Delta})}$$
$$\Rightarrow \hat{\Delta} = \log\left(\frac{\hat{n}_h}{\hat{n}_l} M\right). \tag{A8}$$

Consider introducing a pseudocount for unobserved states, $\lambda$. By observing $\lambda$ more samples, the fraction of these observed states, $\lambda/(M + \lambda)$, will be excited states according to our model. Assuming one additional pseudosample on the unobserved-state side, setting $\lambda = 1$ recovers Eq. (A8).

The effect of this regularization is to learn an energy gap, $\hat{\Delta}$, that is too low with respect to the true energy gap, $\Delta$; thus, necessitating the need to lower sampling temperature $\tau$. Russ *et al.* noted in Ref. [1] that the introduction of regularization into the fitting procedure compensates for limited training data and contributes to the deviations of the fit model's samples toward higher energies. We can see from our regularization used in Eq. (A8) that its effect is the same: Compensating for limited data, inducing generative samples with too high energies via $\hat{\Delta}$.

In Ref. [1], the authors used $L2$ regularization on the model parameters themselves, instead of a pseudocount. The effect is the same, however, smaller parameters lead to higher energies.

Importantly, though regularization contributes to learning smaller parameters, be they $\hat{\Delta}$ or $\hat{J}$ in Ref. [1], it is not necessary to induce the temperature-tuning phenomenon. This is exemplified in Sec. II C, where training proceeds *without* any regularization of the model parameters, yet the fit model still requires *post hoc* tuning to achieve optimal generative performance. Though regularization can exacerbate the tendency of fit models to overestimate the probability mass on excited energy states (cf. Ref. [15] for further exploration of this tendency), the bias inherent in the fit model is also due to bias in the choice of objective function and limited amount of training data, as explored in Appendix D 1.

Algorithm I gives the fitting procedure for finding $\hat{\Delta}$ and $\hat{\mathbf{L}}$ given $\mathbf{p}_{\mathcal{D}}$.

We may further simplify fitting the model by reordering the states in $\mathbf{p}_{\mathrm{data}}$, $i \to k$, such that $p_{\mathrm{data},k} \geqslant p_{\mathrm{data},k+1}$ for $1 \leqslant k \leqslant N_s - 1$, and defining the quantity

$$\mathcal{G}(\ell) := \sum_{k=1}^{\ell} p_{\mathrm{data},k}$$
$$= 1 - \mathbf{L} \cdot \mathbf{p}_{\mathrm{data}},$$

where we note that $k = 1, 2, \ldots, \ell$ index the ground states and therefore $\ell = n_l$. Substituting Eq. (A4) into Eq. (A2) and taking $\tilde{\mathbf{L}} \cdot \mathbf{p}_{\mathrm{data}} \to 1 - \mathcal{G}(\tilde{\ell})$, we may define the objective function

$$\mathcal{L}(\tilde{\ell}) = -(1 - \mathcal{G}(\tilde{\ell}))\log(1 - \mathcal{G}(\tilde{\ell})) - \mathcal{G}(\tilde{\ell})\log\mathcal{G}(\tilde{\ell})$$
$$+ (1 - \mathcal{G}(\tilde{\ell}))\log\left(1 - \frac{\tilde{\ell}}{N_s}\right) + \mathcal{G}(\tilde{\ell})\log\frac{\tilde{\ell}}{N_s}, \tag{A9}$$

which is the same as Eq. (A2) up to a constant that does not depend on $\tilde{\ell}$.

Equation (A9) is optimized over $\tilde{\ell}$, and $\hat{\ell}$ is used to find $\hat{\Delta}$ and $\hat{\mathbf{L}}$.

### 3. Exact expressions for $\tau^*$ and $\tau'$

The toy model is tractable such that we may take derivatives with respect to $\tau$ of $D_{\mathrm{KL}}(\hat{\mathbf{q}}_\tau||\mathbf{p})$ and $D_{\mathrm{KL}}(\mathbf{p}||\hat{\mathbf{q}}_\tau)$ directly and using this to find $\tau^*$ and $\tau'$. The key is to break up the sum over all states into four separate contributions from each of the possible combinations of level assignments per state: (1) $\mathbf{L} = 0$, $\hat{\mathbf{L}} = 0$ ("found low-energy states"), (2) $\mathbf{L} = 0$, $\hat{\mathbf{L}} = 1$ ("missed low"), (3) $\mathbf{L} = 1$, $\hat{\mathbf{L}} = 1$ ("found high"), and (4) $\mathbf{L} = 1$, $\hat{\mathbf{L}} = 0$ ("missed high").

Noting that the number of found high-energy states is $\mathbf{L} \cdot \hat{\mathbf{L}}$ and taking care to break up the sum over states accordingly, we can write

$$D_{\mathrm{KL}}(\hat{\mathbf{q}}_\tau||\mathbf{p})$$
$$= \sum_i \hat{q}_{\tau,i} \log\frac{\hat{q}_{\tau,i}}{p_i}$$
$$= \log\left(\frac{Z}{\hat{Z}(\tau)}\right) - \sum_i \frac{\exp(-\hat{\Delta}\hat{L}_i/\tau)}{\hat{Z}(\tau)}\left(\frac{\hat{\Delta}\hat{L}_i}{\tau} - \Delta L_i\right)$$
$$= \log\left(\frac{Z}{\hat{Z}(\tau)}\right) - \frac{\exp(-\hat{\Delta}/\tau)}{\hat{Z}(\tau)}\left[(\hat{n}_h - \hat{\mathbf{L}}\cdot\mathbf{L})\frac{\hat{\Delta}}{\tau}\right.$$
$$\left. + \hat{\mathbf{L}}\cdot\mathbf{L}\left(\frac{\hat{\Delta}}{\tau} - \Delta\right)\right] + \frac{1}{\hat{Z}(\tau)}(n_h - \hat{\mathbf{L}}\cdot\mathbf{L})\Delta. \tag{A10}$$

ALGORITHM 1. Find $\hat{\Delta}$, $\hat{\mathbf{L}}$.

```
given p_D = (p_{D,1}, p_{D,2}, ..., p_{D,N}), init containers:

losses=LIST[LENGTH=N]                     ▷ ordered list for losses at each training iteration
deltas=LIST[LENGTH=N]                     ▷ for energy gap estimates at each iter
g={ }                                     ▷ collection of indices of states assigned ground energy level. init as empty
e={1, 2, ..., N}                          ▷ indices of states assigned excited energy level. init all states as excited
R={p_{D,1}, p_{D,2}, ..., p_{D,N}}        ▷ collection of empirical frequencies of states
all_e=LIST[LENGTH=N]
all_g=LIST[LENGTH=N]                      ▷ to store state-assignment collections at each learning step
L=LIST[LENGTH=N]; L[i]=1 for i=1, 2, ..., N   ▷ state assignment vector. init all states as excited

for t in 1 to N do
    find current largest p_{D,i} in R and assign corresponding state to ground energy level
    p̄ = LARGEST(R)
    ī = INDEXOF(p̄)
    L[ī] = 0
    calculate estimate of energy gap and associated loss
    deltas[t]=GETENERGYGAP(L, p_D)        ▷ using Eq. (A4).
    losses[t]=GETLOSS(deltas[t], L)       ▷ using Eq. (A2).
    update and record current set of states at each energy level
    e=e\ī
    g = g∪{ī}
    all_e[t]=e
    all_g[t]=g
    remove largest p_{D,i} from R
    R=R\p̄

return L and deltas[t] corresponding to minimum of losses
t̂ = INDEXOFMIN(losses)
L[i]=1 ∀ i ∈ all_e[t̂]
L[i]=0 ∀ i ∈ all_g[t̂]
RETURN(deltas[t̂],L)
```

Finding the stationary point with respect to $\tau$ yields

$$\tau^* = \frac{\hat{n}_h}{\hat{\mathbf{L}} \cdot \mathbf{L} + \frac{\hat{n}_h}{\hat{n}_l}(\hat{\mathbf{L}} \cdot \mathbf{L} - n_h)} \frac{\hat{\Delta}}{\Delta}. \tag{A11}$$

Similarly, we have

$$\begin{aligned} D_{\mathrm{KL}}(\mathbf{p}||\hat{\mathbf{q}}_\tau) &= \log\left(\frac{\hat{Z}(\tau)}{Z}\right) + \frac{1}{Z}(\hat{n}_h - \hat{\mathbf{L}} \cdot \mathbf{L})\frac{\hat{\Delta}}{\tau} \\ &\quad - \frac{\exp(-\Delta)}{Z}\left[(n_h - \hat{\mathbf{L}} \cdot \mathbf{L})\Delta - \hat{\mathbf{L}} \cdot \mathbf{L}\left(\frac{\hat{\Delta}}{\tau} - \Delta\right)\right] \end{aligned} \tag{A12}$$

and

$$\begin{aligned} \tau' &= \frac{\hat{\Delta}}{\Delta - \log x} \\ x &= \frac{\hat{n}_l(\hat{\mathbf{L}} \cdot \mathbf{L} + e^{\Delta}(n_h - \hat{\mathbf{L}} \cdot \mathbf{L} + n_l + \hat{n}_l))}{\hat{n}_h(\hat{n}_l + (n_h - \hat{\mathbf{L}} \cdot \mathbf{L})(e^{-\Delta} - 1))}. \end{aligned} \tag{A13}$$

Equations (A11) and (A13) were used to calculate $\tau^*$ and $\tau'$ for each experiment in Figs. 2(d)–2(h) and Figs. 4(c) and 4(f).

### 4. Limit under which temperature tuning is required

When the energy gap, $\Delta$, in the ground truth is large, high-energy/excited states are rarely sampled. However, in the rare case where they happen to be sampled the empirical frequency on such an excited state, which is $1/M$, grossly overestimates the true frequency with which that given state should occur. We may write this as

$$\frac{1}{M} \gg \frac{\exp(-\Delta)}{Z} \approx \frac{\exp(-\Delta)}{n_l}, \tag{A14}$$

where we note that in the large energy gap regime, the partition function may be approximated as $Z = n_l + n_h \exp(-\Delta) \approx n_l$.

Rearranging Eq. (A14), we obtain the condition on $\Delta$ under which high-energy states will not be well sampled, and therefore overfit to by the model:

$$\Delta \gg \log(M/n_l). \tag{A15}$$

We also note that this defines the undersampled regime as

$$M \ll n_l \exp(\Delta). \tag{A16}$$

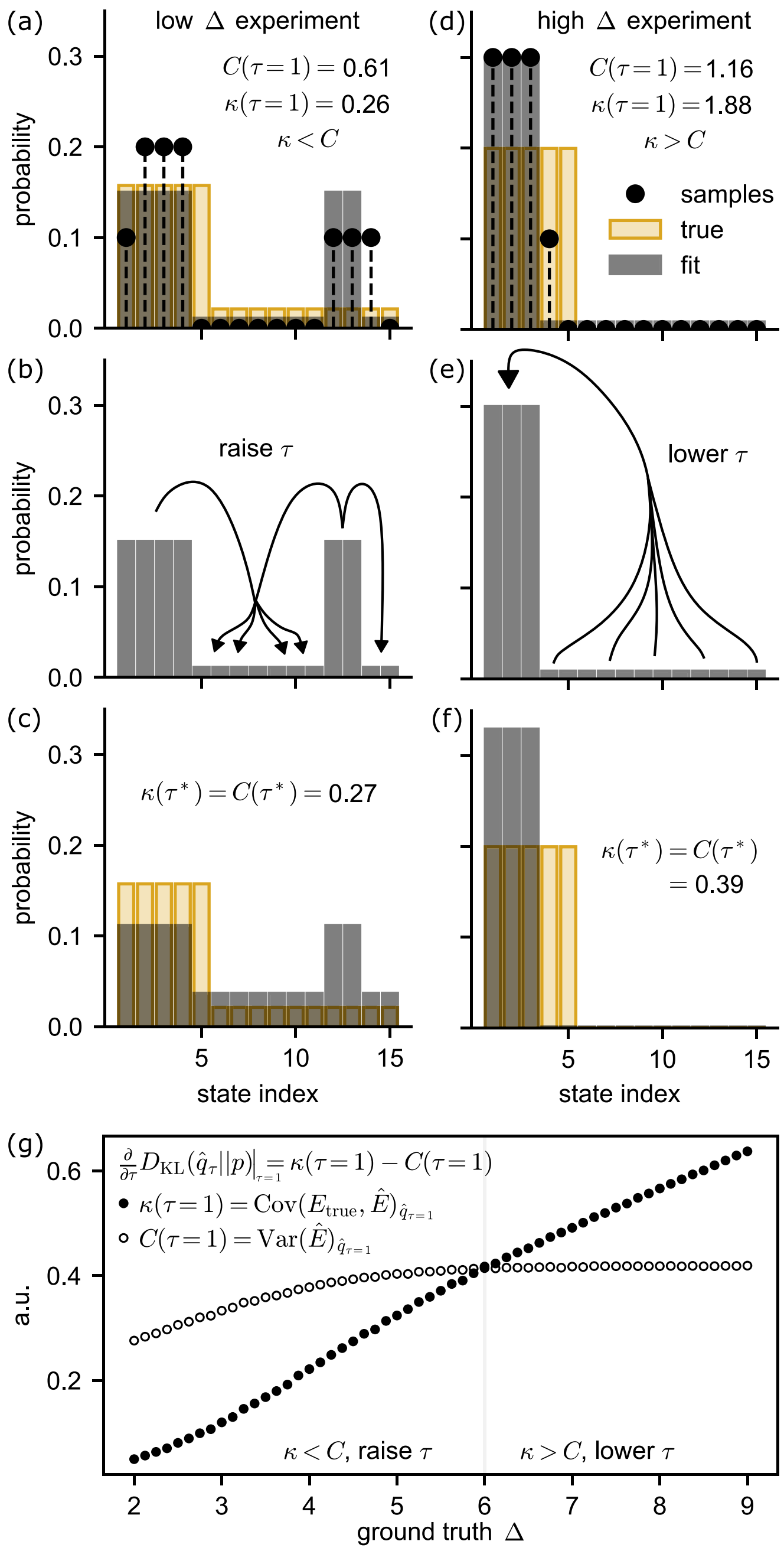

FIG. 4. $\kappa$ and $C$ [Eqs. (C2) and (C3)] determine whether to raise or lower $\tau$ in order to improve generative performance. (a)–(f) Experiments on the illustrative toy model for five low-energy states and ten high-energy states, with models fit to ten training samples. (a)–(c) One experiment for ground truth $\Delta = 2$. Low training data cause erroneous assignment of excited states as ground states in the model (a); weak correlation of model with true distribution, $\kappa < C$, makes it advantageous to raise $\tau$ [panels (b) and (c)]. (d)–(f) One experiment for $\Delta = 7$. Strong correlation of model with true distribution, $\kappa > C$ in panel (d), makes it advantageous to lower $\tau$ [panels (e) and (f)]. In panel (g), each point represents an average of 200 replicates of experiments done for several values of $\Delta$, fixed at 80 training samples. The illustrative toy model contains 20 low-energy states and 80 high-energy states as in Figs. 2(g) and 2(h).

## APPENDIX B: ISING MODEL DISTRIBUTION

### 1. Training from samples

The model is trained until the relative change in the negative log-likelihood goes below $10^{-5}$, that is,

$$\frac{|\mathcal{L}^{\{t\}} - \mathcal{L}^{\{t-1\}}|}{|\mathcal{L}^{\{t-1\}}|} < 10^{-5},$$

where $t$ indexes the training iteration of the gradient descent. Note that gradient updates, which are defined as

$$\tilde{J}_{ij}^{(t+1)} = \tilde{J}_{ij}^{(t)} + \eta(\langle v_i v_j\rangle_{\mathcal{D}_T} - \langle v_i v_j\rangle_{\mathcal{M}_{(t)}}),$$

with learning rate $\eta$, can be calculated exactly since expectation values with respect to the learned model at iteration $t$, $\langle v_i v_j\rangle_{\mathcal{M}_{(t)}}$, can be taken over all $2^{16}$ states stored in memory, and do not require a sampling approximation.

For all experiments, the learning rate, $\eta$, was set to 0.01. Fit parameters were initialized $\tilde{J}_{ij} \sim \mathcal{N}(0, 1)$ for all $i$ and $j$.

### 2. Calculating $\tau^*$ and $\tau'$

For the values $\tau^*$ and $\tau'$ reported in Figs. 3(e)–3(i), exact expressions are not available as in the case of the illustrative toy model. For each model fit, $\hat{q}$, a vector of probabilities for all $2^{16}$ is generated for each value of $\tau$ (swept in the interval [0.2, 5]) and compared to the corresponding $p_T$ (which generated the training data) via the forward and reversed $D_{\mathrm{KL}}$. Resulting $D_{\mathrm{KL}}$ versus $\tau$ plots are fit with a spline function of degree 4, with exact interpolation (0 smoothing), and zero boundary conditions. $\tau^*$ and $\tau'$ are then calculated from the resulting curves.

In Figs. 3(h) and 3(i), ten replicates of experiments are done for each value of $M$ and $T$, and $\tau^*$, $\tau'$ are calculated as above, then averaged. The scaled color image represents the mean over these ten replicates.

## APPENDIX C: WHEN TO RAISE VERSUS WHEN TO LOWER SAMPLING TEMPERATURE $\tau$

To understand the conditions under which $\tau$ needs to be raised or lowered, we must understand the sensitivities of the reversed $D_{\mathrm{KL}}$ to changes in $\tau$—how much it is punished by removing probability from areas in the support with high mass, and how much entropy is gained as this mass is redistributed. To this end, we define the following quantities:

$$\frac{\partial D}{\partial \tau} := \frac{\partial}{\partial \tau} D_{\mathrm{KL}}(\hat{q}_\tau || p) = \frac{1}{\tau}(\kappa(\tau) - C(\tau)), \tag{C1}$$

$$\kappa(\tau) := \tau \frac{H[\hat{q}_\tau, p]}{\partial \tau} = \frac{1}{T_p \tau}\mathrm{Cov}(E_{\mathrm{true}}, \hat{E})_{\hat{q}_\tau}, \tag{C2}$$

$$C(\tau) := \tau \frac{\partial H[\hat{q}_\tau]}{\partial \tau} = \frac{1}{\tau^2}\mathrm{Var}(\hat{E})_{\hat{q}_\tau}, \tag{C3}$$

where $T_p$ is a temperature parameter for the ground truth and where we identify the latter quantity as the heat capacity from traditional statistical physics.

Equation (C1) indicates whether $\tau$ should be raised or lowered, as it determines the gradient along which changes in $\tau$ lead to decreases in the reversed $D_{\mathrm{KL}}$.

Furthermore, as we also see from Eq. (C1) the sign of $\partial D/\partial\tau$ is controlled by the difference between $\kappa$ and $C$. These two quantities may be thought of as susceptibilities parametrized by $\tau$. $\kappa$ measures the propensity for probability mass to come off of true low-energy states if $\tau$ is raised and is proportional to the covariance of true and fit energy functions. If $\hat{E}$ is strongly correlated with $E_{\mathrm{true}}$, there is a stronger penalty for raising $\tau$ and taking mass off of those true low $E$ states. $C$ measures propensity for probability mass to cover more states as $\tau$ is raised.

#### 1. Example from illustrative model

Figure 4 shows the two scenarios of the illustrative toy model for 15 states in which $\tau$ is adjusted (5 low-energy, 10 high, and $T_p = 1$ without loss of generality). Panels (a)–(c) contain one experiment for a ground truth $\Delta = 2$. In panel (a), some true excited states are assigned as model ground states, leading to poor correlation of $\hat{E}$ with $E_{\mathrm{true}}$ and $\kappa < C$. Panel (b) shows $\tau$ being raised to mitigate this mismatch and in panel (c) we can see that at optimal $\tau^*$ the two are equal and therefore, $\partial D/\partial\tau = 0$. Panels (d)–(f) show an experiment for ground truth $\Delta = 7$. Since $\kappa > C$, the penalty for taking mass off of true low-energy states is too great relative to the propensity to cover more of state space, and $\tau$ should therefore be lowered.

Figure 4(g) shows experiments for many values of ground truth $\Delta$, with 200 replicates each, for 20 low-energy states and 80 high-energy states, corresponding to the same system used in Figs. 2(g) and 2(h), with 80 training samples for each replicate. We can see that on average, in this low sample regime, low true energy gaps produce the need to raise $\tau$ as $\kappa < C$, and high-energy gaps necessitate the need to lower $\tau$ because $\kappa > C$.

#### 2. For nearest-neighbor Ising experiments

Figure 5 depicts the difference in conditions under which $\tau$ should be raised or lowered as dictated by the reversed $D_{\mathrm{KL}}$ for experiments on the nearest-neighbor Ising distribution. For an experiment done at low $M$ and *low* $T$ in panels (a) and (b), we see the typical strong contribution from the excited states. The positive value of the derivative with respect to $\tau$ of the reversed $D_{\mathrm{KL}}$ indicates it is advantageous to lower $\tau$. For an experiment done at low $M$ and *high* $T$ in panels (c) and (d), a similar strong contribution from excited states to the reversed $D_{\mathrm{KL}}$ dominates. However, the derivative is negative, and therefore implies $\tau$ should be raised.

The difference between $\kappa$ and $C$—which track the strength of the covariance of $\hat{E}$ with $E_{\mathrm{true}}$ (the model's susceptibility to move probability mass on/off true low-energy states) and the model's energy variance (the model's susceptibility to spread out probability mass over state space in general)—controls the sign of $\partial D/\partial\tau$, and therefore the direction in which $\tau$ should be changed [cf. Fig. 4(g)]. This is clearly seen in Fig. 5(e). At low temperatures $\kappa > C$, and $\tau$ must be lowered, while at high $T$, $\kappa < C$ and $\tau$ should be raised.

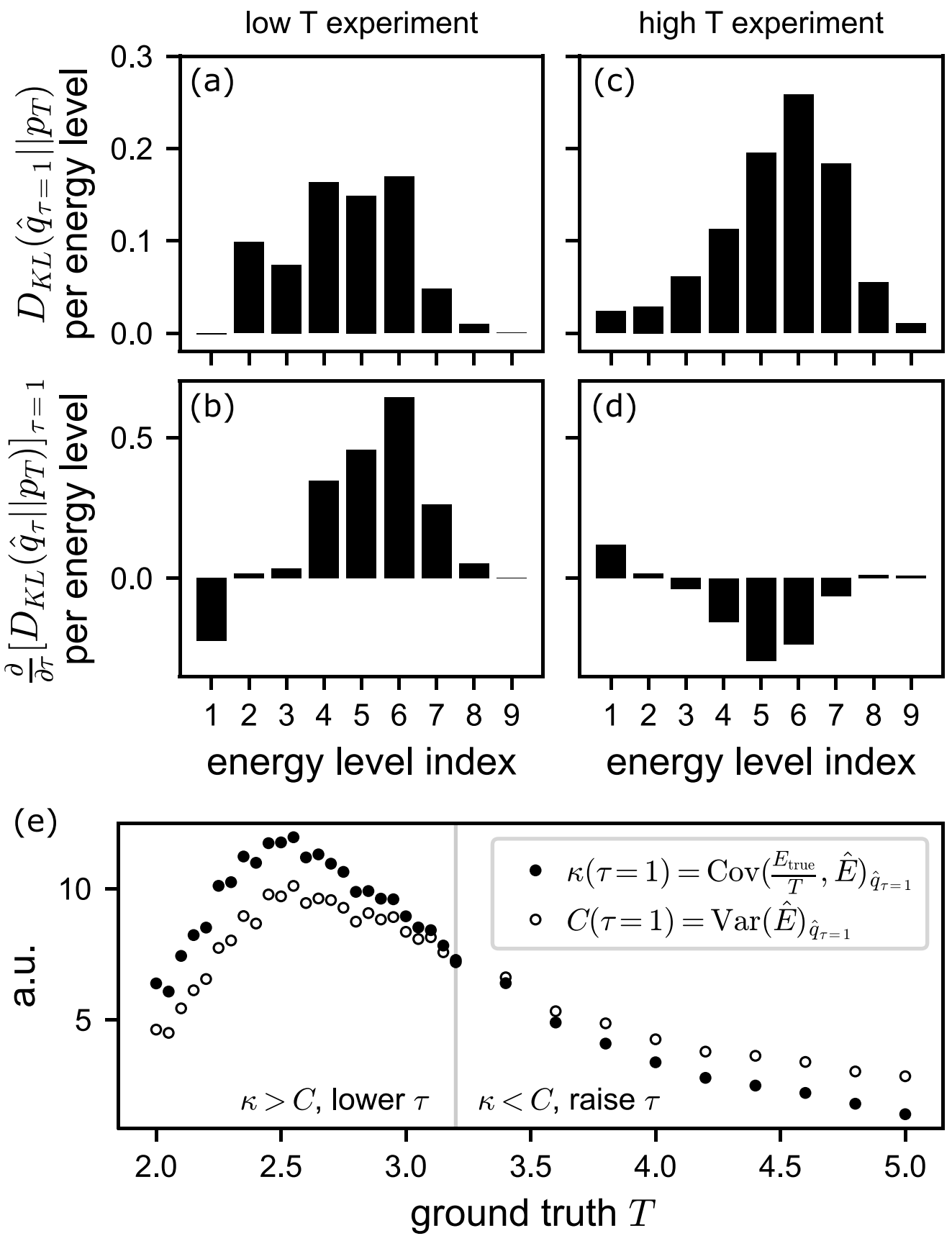

FIG. 5. Per-energy-level breakdown of the reversed $D_{\mathrm{KL}}$ and its derivative with respect to $\tau$ reveals what dictates the need to raise and lower $\tau$. (a), (b) Results of an experiment done on the $4 \times 4$ nearest-neighbor Ising distribution at $T = 2.3$ and $\hat{q}$ trained on $M = 54$ samples. In panel (a), contributions to the reversed $D_{\mathrm{KL}}$ are shown for the first nine excited energy levels. In panel (b), contributions to its derivative with respect to $\tau$ evaluated at $\tau = 1$ are positive, indicative of a need to lower $\tau$. (c), (d) Results of an experiment done at $T = 4$ for $M = 54$. Strong contributions to reversed $D_{\mathrm{KL}}$ also come from excited states [panel (c)]; however, negative contributions dominate the derivative [panel (d)], revealing a need to raise $\tau$. (e) Many experiments done on the $4 \times 4$ Ising distribution at various ground truth $T$. Each point represents an average over ten experiments done with 54 training data each. For low $T$, the need to lower $\tau$ is necessitated by a strong correlation to the true energy function relative to the model's energy variance; $\kappa > C$, corresponding to a positive value of $\frac{\partial}{\partial\tau}D_{\mathrm{KL}}(\hat{q}_\tau||p_T)|_{\tau=1}$. For high $T$, the intramodel variance dominates, and probability mass can spread out over state space faster than it comes off of true low-energy states, i.e., $\kappa < C$ and $\tau$ should be raised.

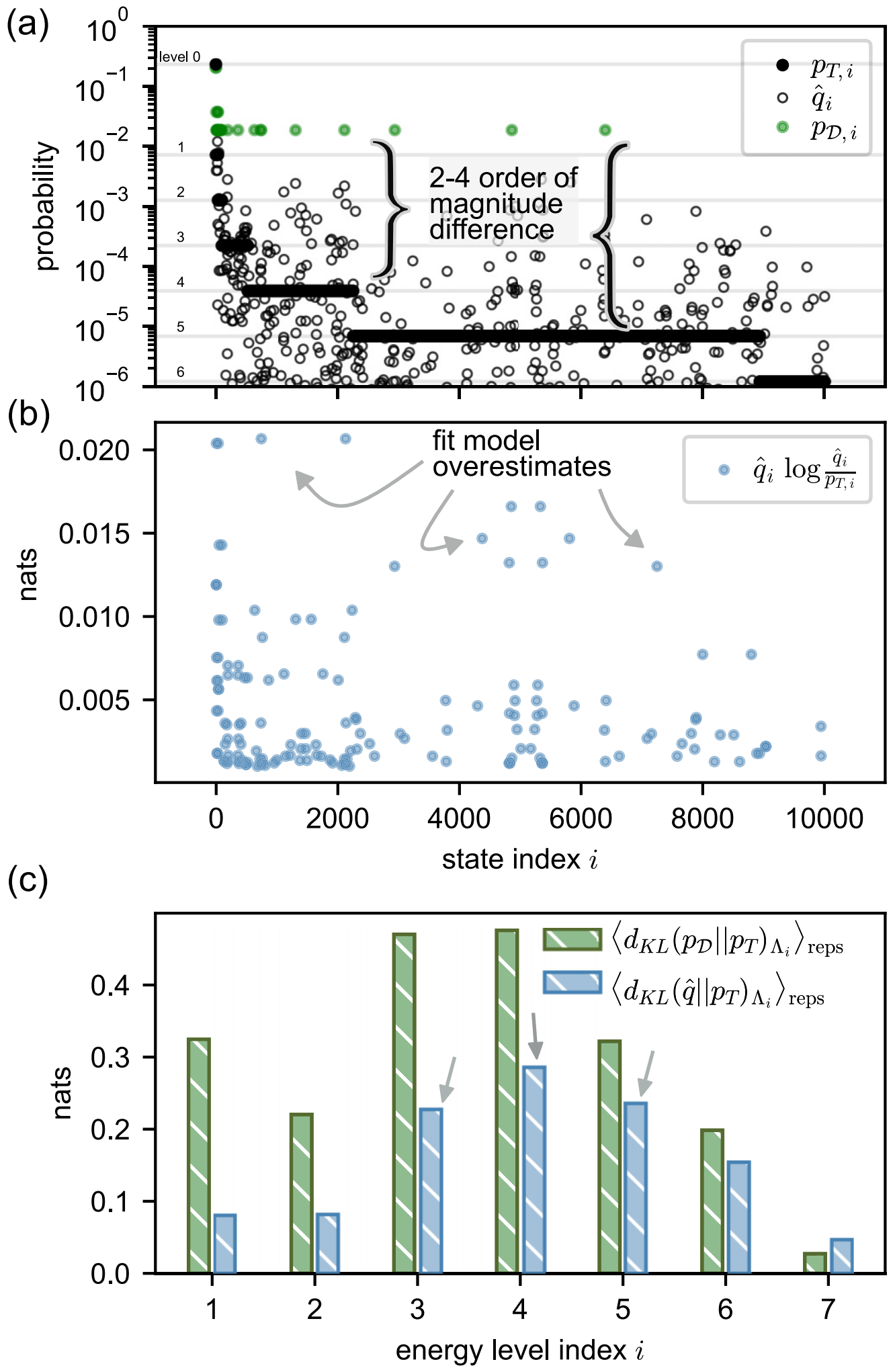

FIG. 6. Bias introduced by the empirical approximation to $D_{\rm KL}(p||q)$ with $D_{\rm KL}(p_{\mathcal{D}}||q)$. (a) Enumerated probability masses for the first 10 000 states of the ground and first six excited energy levels of a $4 \times 4$ nearest-neighbor Ising model, for a ground truth $p_T$ at $T = 2.3$. $p_{\mathcal{D}}$ is the empirical distribution formed by $M = 54$ samples from $p_T$. $\hat{q}$ is the maximum likelihood estimate found via Eq. (8). Note that for excited energy states we have $p_{\mathcal{D}} \gg p_T$ ($\hat{q} \gg p_T$ for many, though not all, states, as well). $\hat{q}$ is subsampled to 1000 randomly chosen states for clarity. (b) $\hat{q}_i \log \frac{\hat{q}_i}{p_{T,i}}$ is the contribution to the reversed $D_{\rm KL}$ per state (values below $10^{-3}$ omitted for clarity). (c) $D_{\rm KL}$'s decomposed according to Eq. (13) and averaged over 50 replicates. $D_{\rm KL}(p_{\mathcal{D}}||p_T)$ is generically large for higher excited states, and consequently so is $D_{\rm KL}(\hat{q}||p_T)$.

## APPENDIX D: OBJECTIVE FUNCTION MISALIGNMENT WITH GENERATIVE PERFORMANCE GOALS

### 1. Bias introduced by an empirical approximation of the objective function

Without the true distribution, the objective function must be approximated as in Eq. (3), $D_{\rm KL}(p||q) \approx D_{\rm KL}(p_{\rm data}||q)$. What kind of bias might this introduce to learning?

Figure 6(a) depicts the probabilities of the first 10 000 states of a $4 \times 4$ nearest-neighbor Ising model, the associated empirical distribution $p_{\mathcal{D}}$ from training data consisting of $M = 54$ samples from a ground truth distribution at $T = 2.3$, and the associated fit model $\hat{q}$. Note that the minimum value $p_{\mathcal{D}}(v)$ can take is $1/M$, which is well above the probability of higher excited states. Concomitantly, $\hat{q}$, representing our best guess of $p_T$, overestimates many of these excited states, harming the reversed $D_{\rm KL}$ [Fig. 6(b)].

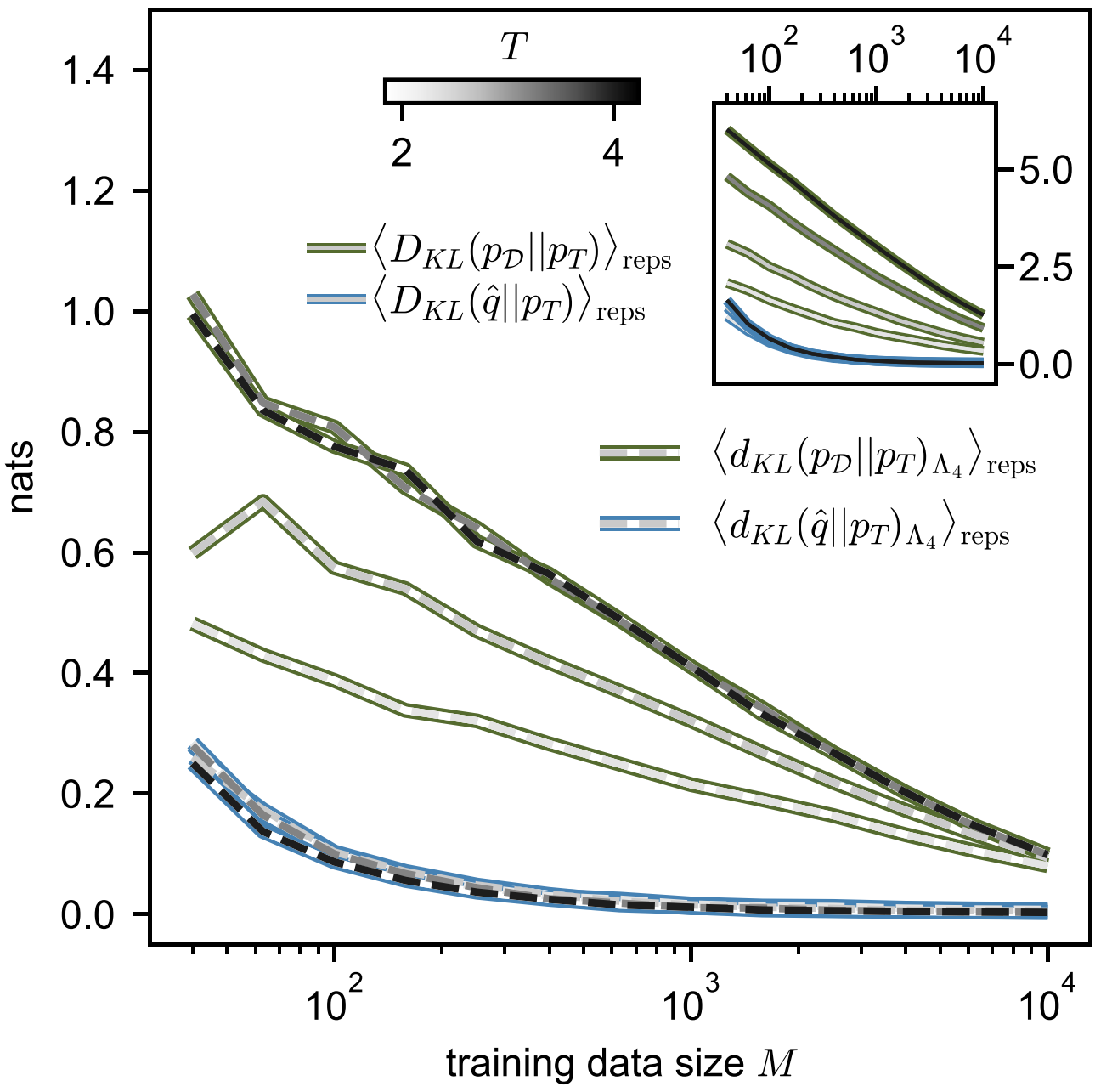

FIG. 7. Mean values of $d_{\rm KL}(p_{\mathcal{D}}||p_T)_{\Lambda_4}$ and $d_{\rm KL}(\hat{q}||p_T)_{\Lambda_4}$ over 50 replicates of experiments for several values of $M$ and $T$. The value of $T$ is denoted by the shade of black of each line. Each $D_{\rm KL}$ decreases as $M$ increases, though $d_{\rm KL}(p_{\mathcal{D}}||p_T)_{\Lambda_4} > d_{\rm KL}(\hat{q}||p_T)_{\Lambda_4}$ always. (Inset) This bound is obeyed when considering the total value of each $D_{\rm KL}$. Means with respect to replicates are calculated as in Fig. 6.

Figure 6(c) shows values of $D_{\rm KL}(p_{\mathcal{D}}||p_T)$ and $D_{\rm KL}(\hat{q}||p_T)$, each broken down according to their contributions from the first seven excited energy levels, using Eq. (13).

Reported are averages over 50 replicates of an experiment with $T = 2.3$, $M = 40$ for a $4 \times 4$ Ising distribution. To ensure adequate comparison across the $P = 50$ replicates, we calculate means reported above as $\langle D_{\rm KL}(f||g)\rangle \times \frac{1}{P}\sum_{k=1}^{P}\frac{[d_{\rm KL}(f||g)_{\Lambda_i}]_{(k)}}{[D_{\rm KL}(f||g)]_{(k)}}$, where the first term is the mean of the total $D_{\rm KL}$ and the second term is the mean fractional contribution from each energy level to the total $D_{\rm KL}$.

We see that, on average, for these excited energy levels we have

$$d_{\rm KL}(p_{\mathcal{D}}||p_T)_{\Lambda_i} \gtrsim d_{\rm KL}(\hat{q}||p_T)_{\Lambda_i},$$

with excited states 3–5 contributing most of this systematic overestimation, and therefore, adversely affecting generative performance (see Appendix D 2 for further discussion on this bound).

The excited energy levels in the empirical distribution, $p_{\mathcal{D},e}$, are systematically overrepresented with respect to $p_{T,e}$, and $\hat{q}_e$ are overestimated accordingly.

### 2. Upper bounds on $D_{\mathrm{KL}}(\hat{q}||p)$

$D_{\mathrm{KL}}(\hat{q}||p_T)$ is bounded from above by $D_{\mathrm{KL}}(p_{\mathcal{D}}||p_T)$ per energy level, which makes intuitive sense—we would expect our best estimate of the distribution to be no worse than the data itself.

In information geometry, a standard identity (Pythagorean relation for the KL divergence) relates the maximum likelihood estimate of a model to data taken from a distribution (that is, of $\hat{q}$ to $p_{\mathcal{D}}$ taken from $p$) [58]

$$D_{\mathrm{KL}}(p_{\mathcal{D}}||p) = D_{\mathrm{KL}}(p_{\mathcal{D}}||\hat{q}) + D_{\mathrm{KL}}(\hat{q}||p). \tag{D1}$$

This is true when $\hat{q}$ is in the exponential family of distributions and the support of $p_{\mathcal{D}}$ and $\hat{q}$ are the same. The same support is shared by both distributions only when $p_{\mathcal{D}}$ is well sampled, that is, every state is sampled at least once. This is especially rare for low $M$ and low $T$, as many states are not sampled. However, Eq. (D1), implies the following bound:

$$D_{\mathrm{KL}}(p_{\mathcal{D}}||p) > D_{\mathrm{KL}}(\hat{q}||p) \tag{D2}$$

and furthermore, we expect this bound to hold in the limiting behavior, $M \to \infty$. We find empirically that this bound is mostly obeyed on a per-energy-level basis, on average, for low $M$ experiments. In Fig. 7, many such experiments are conducted for various values of $T$ and $M$ on the $4 \times 4$ Ising model. We see that for energy level 4, $d_{\mathrm{KL}}(p_{\mathcal{D}}||p_T)_{\Lambda_4} > d_{\mathrm{KL}}(\hat{q}||p_T)_{\Lambda_4}$, even as $M$ is quite low, and for various values of $T$.

[1] W. P. Russ, M. Figliuzzi, C. Stocker, P. Barrat-Charlaix, M. Socolich, P. Kast, D. Hilvert, R. Monasson, S. Cocco, M. Weigt, and R. Ranganathan, An evolution-based model for designing chorismate mutase enzymes, Science **369**, 440 (2020).

[2] G. Hinton, O. Vinyals, and J. Dean, Distilling the knowledge in a neural network, arXiv:1503.02531.

[3] Q. Qi, J. Lyu, K.-S. Chan, E.-W. Bai, and T. Yang, Stochastic constrained DRO with a complexity independent of sample size, Trans. Mach. Learn. Res. **2023**, 1 (2023).

[4] Z.-H. Qiu, Q. Hu, Z. Yuan, D. Zhou, L. Zhang, and T. Yang, Not all semantics are created equal: Contrastive self-supervised learning with automatic temperature individualization, in *Proceedings of the 40th International Conference on Machine Learning, ICML'23* (JMLR.org, Honolulu, Hawaii, 2023), Vol. 202, pp. 28389–28421.

[5] C. Guo, G. Pleiss, Y. Sun, and K. Q. Weinberger, On calibration of modern neural networks, in *Proceedings of the 34th International Conference on Machine Learning* (PMLR, Sydney, Australia, 2017), pp. 1321–1330.

[6] F. Wenzel, K. Roth, B. Veeling, J. Swiatkowski, L. Tran, S. Mandt, J. Snoek, T. Salimans, R. Jenatton, and S. Nowozin, How good is the Bayes posterior in deep neural networks really? in *Proceedings of the 37th International Conference on Machine Learning* (PMLR, 2020), pp. 10248–10259.

[7] Y. Zhang, Y.-S. Wu, L. A. Ortega, and A. R. Masegosa, The cold posterior effect indicates underfitting, and cold posteriors represent a fully Bayesian method to mitigate it, Trans. Mach. Learn. Res. **2024**, 1 (2024).

[8] E. Zhang, V. Zhu, N. Saphra, A. Kleiman, B. L. Edelman, M. Tambe, S. Kakade, and E. Malach, Transcendence: Generative models can outperform the experts that train them, Adv. Neural Inf. Process. Syst. **37**, 86985 (2024).

[9] H. Ramsauer, B. Schäfl, J. Lehner, P. Seidl, M. Widrich, L. Gruber, M. Holzleitner, T. Adler, D. Kreil, M. K. Kopp, G. Klambauer, J. Brandstetter, and S. Hochreiter, Hopfield networks is all you need, in *International Conference on Learning Representations* (OpenReview.net, 2021).

[10] D. Firth, Bias reduction of maximum likelihood estimates, Biometrika **80**, 27 (1993).

[11] M. B. Kloucek, T. Machon, S. Kajimura, C. P. Royall, N. Masuda, and F. Turci, Biases in inverse Ising estimates of near-critical behavior, Phys. Rev. E **108**, 014109 (2023).

[12] B. Bordelon, A. Canatar, and C. Pehlevan, Spectrum dependent learning curves in kernel regression and wide neural networks, in *Proceedings of the 37th International Conference on Machine Learning* (PMLR, 2020), pp. 1024–1034.

[13] B. Sorscher, R. Geirhos, S. Shekhar, S. Ganguli, and A. Morcos, Beyond neural scaling laws: Beating power law scaling via data pruning, Adv. Neural Inf. Process. Syst. **35**, 19523 (2022).

[14] Y. Kleeorin, W. P. Russ, O. Rivoire, and R. Ranganathan, Undersampling and the inference of coevolution in proteins, Cell Syst. **14**, 210 (2023).

[15] P. W. Fields, V. Ngampruetikorn, R. Ranganathan, D. J. Schwab, and S. Palmer, Understanding energy-based modeling of proteins via an empirically motivated minimal ground truth model, in *1st Workshop on the Synergy of Scientific and Machine Learning Modeling, ICML 2023* (Honolulu, Hawaii, 2023).

[16] J. M. Smith, Natural selection and the concept of a protein space, Nature (London) **225**, 563 (1970).

[17] Y. Ishida, Y. Ichikawa, A. Dote, T. Miyazawa, and K. Hukushima, Ratio divergence learning using target energy in restricted Boltzmann machines: Beyond Kullback-Leibler divergence learning, Phys. Rev. E **112**, 045306 (2025).

[18] F. Huszár, How (not) to train your generative model: Scheduled sampling, likelihood, adversary? arXiv:1511.05101.

[19] P. Mehta, M. Bukov, C.-H. Wang, A. G. R. Day, C. Richardson, C. K. Fisher, and D. J. Schwab, A high-bias, low-variance introduction to machine learning for physicists, Phys. Rep. **810**, 1 (2019).

[20] I. J. Goodfellow, J. Pouget-Abadie, M. Mirza, B. Xu, D. Warde-Farley, S. Ozair, A. Courville, and Y. Bengio, Generative adversarial nets, in *Advances in Neural Information Processing Systems* (Curran Associates, Inc., Red Hook, NY, 2014), Vol. 27.

[21] C. K. Fisher, A. M. Smith, and J. R. Walsh, Boltzmann encoded adversarial machines, arXiv:1804.08682.

[22] Strictly, $\Delta \to \infty$ is not required. Temperature tuning is required when $\Delta \gg \log(M/n_l)$ (see Appendix A 4 for more details).

[23] It is worth noting that the reversed $D_{\mathrm{KL}}(\mathbf{q}||\mathbf{p})$ is not easily fit with a similar approximation, $D_{\mathrm{KL}}(\mathbf{q}||\mathbf{p}_{\mathrm{data}}) = \sum_i q_i \log \frac{q_i}{p_{\mathrm{data},i}}$, as we typically do not have a tractable empirical estimate for this quantity.

[24] A pseudocount regularization term is introduced to ensure no divergences while fitting (see Appendix A for further details

of the sampling and fitting procedure, as well as a discussion on the contributions of regularization to the temperature-tuning effect).

[25] See Appendix C 2 for more details.

[26] J. B. Ingraham, M. Baranov, Z. Costello, K. W. Barber, W. Wang, A. Ismail, V. Frappier, D. M. Lord, C. Ng-Thow-Hing, E. R. Van Vlack, *et al.*, Illuminating protein space with a programmable generative model, Nature (London) **623**, 1070 (2023).

[27] A. Madani, B. Krause, E. R. Greene, S. Subramanian, B. P. Mohr, J. M. Holton, J. L. Olmos, C. Xiong, Z. Z. Sun, R. Socher, J. S. Fraser, and N. Naik, Large language models generate functional protein sequences across diverse families, Nat. Biotechnol. **41**, 1099 (2023).

[28] T. Hayes, R. Rao, H. Akin, N. J. Sofroniew, D. Oktay, Z. Lin, R. Verkuil, V. Q. Tran, J. Deaton, M. Wiggert, *et al.*, Simulating 500 million years of evolution with a language model, Science **387**, 850 (2025).

[29] E. T. Jaynes, Information theory and statistical mechanics, Phys. Rev. **106**, 620 (1957).

[30] M. Weigt, R. A. White, H. Szurmant, J. A. Hoch, and T. Hwa, Identification of direct residue contacts in protein-protein interaction by message passing, Proc. Natl. Acad. Sci. USA **106**, 67 (2009).

[31] F. Morcos, A. Pagnani, B. Lunt, A. Bertolino, D. S. Marks, C. Sander, R. Zecchina, J. N. Onuchic, T. Hwa, and M. Weigt, Direct-coupling analysis of residue coevolution captures native contacts across many protein families, Proc. Natl. Acad. Sci. USA **108**, E1293 (2011).

[32] T. Mora, A. M. Walczak, W. Bialek, and C. G. Callan, Maximum entropy models for antibody diversity, Proc. Natl. Acad. Sci. USA **107**, 5405 (2010).

[33] J. Harte, *Maximum Entropy and Ecology: A Theory of Abundance, Distribution, and Energetics* (Oxford University Press, Oxford, UK, 2011).

[34] W. Bialek, A. Cavagna, I. Giardina, T. Mora, E. Silvestri, M. Viale, and A. M. Walczak, Statistical mechanics for natural flocks of birds, Proc. Natl. Acad. Sci. USA **109**, 4786 (2012).

[35] T. Mora, A. M. Walczak, L. Del Castello, F. Ginelli, S. Melillo, L. Parisi, M. Viale, A. Cavagna, and I. Giardina, Local equilibrium in bird flocks, Nat. Phys. **12**, 1153 (2016).

[36] T. R. Lezon, J. R. Banavar, M. Cieplak, A. Maritan, and N. V. Fedoroff, Using the principle of entropy maximization to infer genetic interaction networks from gene expression patterns, Proc. Natl. Acad. Sci. USA **103**, 19033 (2006).

[37] J. W. Locasale and A. Wolf-Yadlin, Maximum entropy reconstructions of dynamic signaling networks from quantitative proteomics Data, PLoS One **4**, e6522 (2009).

[38] E. Schneidman, M. J. Berry, R. Segev, and W. Bialek, Weak pairwise correlations imply strongly correlated network states in a neural population, Nature (London) **440**, 1007 (2006).

[39] G. Tkačik, O. Marre, D. Amodei, E. Schneidman, W. Bialek, M. J. Berry, and O. Sporns, Searching for collective behavior in a large network of sensory neurons, PLoS Comput. Biol. **10**, e1003408 (2014).

[40] M. J. Berry II, G. Tkačik, J. Dubuis, O. Marre, and R. A. da Silveira, A simple method for estimating the entropy of neural activity, J. Stat. Mech.: Theory Exp. (2013) P03015.

[41] C. W. Lynn, Q. Yu, R. Pang, S. E. Palmer, and W. Bialek, Exact minimax entropy models of large-scale neuronal activity, Phys. Rev. E **111**, 054411 (2025).

[42] G. Tkačik, T. Mora, O. Marre, D. Amodei, S. E. Palmer, M. J. Berry II, and W. Bialek, Thermodynamics and signatures of criticality in a network of neurons, Proc. Natl. Acad. Sci. USA **112**, 11508 (2015).

[43] D. P. Kingma and M. Welling, An introduction to variational autoencoders, in Found. Trends Mach. Learn. **12**, 307 (2019).

[44] D. Rezende and S. Mohamed, Variational inference with normalizing flows, in *Proceedings of the 32nd International Conference on Machine Learning* (PMLR, Lille, France, 2015), pp. 1530–1538.

[45] J. Sohl-Dickstein, E. Weiss, N. Maheswaranathan, and S. Ganguli, Deep unsupervised learning using nonequilibrium thermodynamics, in *Proceedings of the 32nd International Conference on Machine Learning* (PMLR, Lille, France, 2015), pp. 2256–2265.

[46] A. Vaswani, N. Shazeer, N. Parmar, J. Uszkoreit, L. Jones, A. N. Gomez, Ł. Kaiser, and I. Polosukhin, Attention is all you need, in *Advances in Neural Information Processing Systems* (Curran Associates, Inc., Red Hook, NY, 2017), Vol. 30.

[47] Y. LeCun, S. Chopra, R. Hadsell, M. Ranzato, and F. J. Huang, A tutorial on energy-based learning, *Predicting Structured Data*, edited by G. Bakr, T. Hofmann, B. Schölkopf, A. J. Smola, B. Taskar, and S. V. N. Vishwanathan (MIT Press, Cambridge, MA, 2007).

[48] A. Gu and T. Dao, Mamba: Linear-time sequence modeling with selective state spaces, arXiv:2312.00752.

[49] W. Bialek and R. Ranganathan, Rediscovering the power of pairwise interactions, arXiv:0712.4397.

[50] S. Cocco, C. Feinauer, M. Figliuzzi, R. Monasson, and M. Weigt, Inverse statistical physics of protein sequences: A key issues review, Rep. Prog. Phys. **81**, 032601 (2018).

[51] A. Rives, J. Meier, T. Sercu, S. Goyal, Z. Lin, J. Liu, D. Guo, M. Ott, C. L. Zitnick, J. Ma, and R. Fergus, Biological structure and function emerge from scaling unsupervised learning to 250 million protein sequences, Proc. Natl. Acad. Sci. USA **118**, e2016239118 (2021).

[52] W. Bialek, *Biophysics: Searching for Principles* (Princeton University Press, Princeton, NJ, 2012).

[53] T. Mora and W. Bialek, Are biological systems poised at criticality? J. Stat. Phys. **144**, 268 (2011).

[54] A. De Martino and D. De Martino, An introduction to the maximum entropy approach and its application to inference problems in biology, Heliyon **4**, e00596 (2018).

[55] G. Papamakarios, E. Nalisnick, D. J. Rezende, S. Mohamed, and B. Lakshminarayanan, Normalizing flows for probabilistic modeling and inference, J. Mach. Learn. Res. **22**, 1 (2021).

[56] L. Yang, Z. Zhang, Y. Song, S. Hong, R. Xu, Y. Zhao, W. Zhang, B. Cui, and M.-H. Yang, Diffusion models: A comprehensive survey of methods and applications, ACM Comput. Surv. **56**, 1 (2024).

[57] P. Fields, sepalmer/temp-tune: temperature tuning code v1.0.0 (Version v1.0.0) [Computer software], Zenodo, 2026, https://doi.org/10.5281/zenodo.23333988.

[58] I. Csiszár and P. Shields, *Information Theory and Statistics: A Tutorial*, Foundations and Trends in Communications and Information Theory (Now Publishers Inc., 2004).